\documentclass[11pt]{article}
\usepackage{amsmath}
\usepackage{amsfonts}
\usepackage{amssymb}
\usepackage{graphicx}

\newcommand{\Title}{\bf \sf \huge \noindent}
\newcommand{\Author}{\bf \sf \large \noindent}
\begin{document}
\mbox{  } \vskip 1.2cm

\Title{An exactly solvable phase transition model: generalized
statistics and generalized Bose-Einstein condensation}

\noindent \Huge
------------------------------------------------------
\vskip 0.5cm \Author{Wu-Sheng Dai and Mi Xie}

{\it \noindent \small Department of Physics, Tianjin University,
Tianjin 300072, P.
R. China}\\
{\it \noindent \small LiuHui Center for Applied Mathematics,
Nankai University \& Tianjin
University, Tianjin 300072, P. R. China}\\
{\it \noindent \small Email: \rm daiwusheng@tju.edu.cn,
xiemi@tju.edu.cn}

\thispagestyle{myheadings} \markboth{Preprint}{\underline{J. Stat.
Mech. (2009) P07034}} \vskip 1cm \rm \normalsize

\noindent {\bf Abstract:} In this paper, we present an exactly
solvable phase transition model in which the phase transition is
purely statistically derived. The phase transition in this model
is a generalized Bose-Einstein condensation. The exact expression
of the thermodynamic quantity which can simultaneously describe
both gas phase and condensed phase is solved with the help of the
homogeneous Riemann-Hilbert problem, so one can judge whether
there exists a phase transition and determine the phase transition
point mathematically rigorously. A generalized statistics in which
the maximum occupation numbers of different quantum states can
take on different values is introduced, as a generalization of
Bose-Einstein and Fermi-Dirac statistics.

\noindent

\vskip 0.5cm \noindent Keywords: Rigorous results in statistical
mechanics, Bose Einstein condensation (Theory), Fractional states
(Theory)
\newpage
\vskip 1cm \noindent
---------------------------------------------------------------------------------------------------------------------
\tableofcontents

\vskip 1cm \noindent
---------------------------------------------------------------------------------------------------------------------

\section{Introduction}

A few exactly solvable models play important roles in phase transition theory,
since most, if not all, of our understanding of phase transitions comes from
studying models \cite{es}. In this paper, we present a purely statistically
derived solvable phase transition model. In the model, the exactly solved
thermodynamic quantity can simultaneously describe different phases.
Therefore, whether there is a phase transition can be judged mathematically
rigorously, and the phase transition temperature can be calculated directly by
analyzing the discontinuity in the thermodynamic quantities or their derivatives.

Bose-Einstein condensation (BEC) is the first purely statistically derived
example of a phase transition. The phase transition in the present model is a
generalized Bose-Einstein condensation; in other words, the phase transition
is a BEC type phase transition.

The BEC type phase transition is a sudden change in the microscopic particle
distribution: in the gas phase, no quantum state is macroscopically occupied,
while in the condensed phase, there is a quantum state being occupied by a
macroscopic number of particles. The microscopic particle distribution
determines the macroscopic behavior of a thermodynamic system, or, the
macroscopic behavior reflects the average contribution of all quantum states
in the system. In the condensed phase, the macroscopic behavior of the system
is to a certain extent determined by the single quantum state that is
macroscopically occupied, since the number of particles in such a state is of
the same order of magnitude of the total number of particles. As a macroscopic
manifestation of such a sudden change in the particle distribution, there is a
singularity in the thermodynamic quantity.

We will show that, beyond the Bose case, there are still other systems that
can display BEC type phase transitions.

First, we will introduce a generalized statistics in which different quantum
states have different maximum occupation numbers, and Bose-Einstein and
Fermi-Dirac statistics are its special cases. Especially, we will pay
attention to a special case of the generalized statistics in which at least
one quantum state's maximum occupation number is infinite and show that the
BEC type phase transition may occur in such systems. For example, we will show
that a BEC type phase transition can occur in an any-dimensional ideal gas
obeying the generalized statistics in which the maximum occupation number of
the ground state is infinity, like that in the Bose-Einstein case, and the
maximum occupation number of all other quantum states is $1$, like that in the
Fermi-Dirac case. For comparison, recall that the BEC can occur only in
three-dimensional ideal Bose gases, but cannot occur in one- and
two-dimensional cases.

The mathematical method for solving the model is based on the homogeneous
Riemann-Hilbert problem --- the boundary problem of analytic functions, which
comes from the theory of singular integral equations
\cite{Leonard,Musihailishiweili}.

Moreover, our result also shows that a phase transition occurs only in the
thermodynamic limit, i.e., the total number of particles and the volume must
be infinite. The common proof for this result depends on an assumption that a
finite volume can accommodate at most a finite number of particles, which is,
of course, only valid for non-ideal gases \cite{Huang}. Our result provides an
example that this result holds also for ideal gases.

In section 2, we introduce the generalized statistics. In section 3, we
construct and solve the phase transition model. Discussions and an outlook are
given in section 4.

\section{The generalized statistics}

In this section, we introduce a generalized statistics in which the maximum
occupation number of a quantum state can take on unrestricted integers or
infinity and the maximum occupation numbers of different states may be different.

Let $n_{i}$ be the maximum occupation number of the $i$-th quantum state,
where $n_{i}$ can take on an integer or $\infty$. The grand partition function
is%
\begin{equation}
\Xi\left(  T,V,\mu\right)  =\prod\limits_{i=0}^{\infty}\frac{1-e^{-\beta
\left(  n_{i}+1\right)  \left(  \varepsilon_{i}-\mu\right)  }}{1-e^{-\beta
\left(  \varepsilon_{i}-\mu\right)  }},\label{PF}%
\end{equation}
where $T$ is the temperature, $V$ the volume, $\mu$ the chemical potential,
$\varepsilon_{i}$ the energy of the $i$-th state, and $\beta=1/\left(
k_{B}T\right)  $. Then the equation of state reads%
\begin{align}
\frac{PV}{k_{B}T}  &  =\sum_{i=0}^{\infty}\ln\frac{1-z^{n_{i}+1}e^{-\left(
n_{i}+1\right)  \beta\varepsilon_{i}}}{1-ze^{-\beta\varepsilon_{i}}%
},\label{GPkT}\\
N  &  =\sum_{i=0}^{\infty}\left[  \frac{1}{z^{-1}e^{\beta\varepsilon_{i}}%
-1}-\frac{n_{i}+1}{\left(  z^{-1}e^{\beta\varepsilon_{i}}\right)  ^{n_{i}%
+1}-1}\right]  .\label{GN}%
\end{align}
The equations of state for Bose-Einstein, Fermi-Dirac, and Gentile
\cite{OursAnn,Gentile} cases can be recovered by setting $n_{i}=\infty$,
$n_{i}=1 $, and $n_{i}=n$, respectively.

In this paper, we consider an ideal gas obeying the generalized statistics in
which the maximum occupation number of only one quantum state is $\infty$, but
of all other quantum states is $n$, a given integer, i.e., $n_{k}=\infty$ and
$n_{i}=n$, $\left(  i\neq k\right)  $. The equation of state for such an ideal
gas with the dispersion relation $\varepsilon=p^{s}/\left(  2m\right)  $ in
$\nu$ dimensions can be obtained from equations (\ref{GPkT}) and (\ref{GN}):
\begin{align}
\frac{PV}{k_{B}T}  &  =N_{\lambda}h_{\nu/s+1}\left(  z\right)  -\ln\left[
1-\left(  ze^{-\beta\varepsilon_{k}}\right)  ^{n+1}\right]  ,\\
N  &  =N_{\lambda}h_{\nu/s}\left(  z\right)  +\frac{n+1}{\left(
z^{-1}e^{\beta\varepsilon_{k}}\right)  ^{n+1}-1},\label{Neqofs}%
\end{align}
where $z=e^{\beta\mu}$ is the fugacity, $N_{\lambda}=\frac{2\Gamma\left(
\nu/s\right)  }{s\Gamma\left(  \nu/2\right)  }\frac{V}{\lambda^{\nu}}$, and
$\lambda=\frac{h}{\left(  2\pi^{s/2}mk_{B}T\right)  ^{1/s}}$ is the mean
thermal wavelength. $h_{\sigma}(z)$ can be expressed by the Bose-Einstein
integral, $g_{\sigma}(z)$, as%
\[
h_{\sigma}(z)=g_{\sigma}(z)-\left(  n+1\right)  ^{-\left(  \sigma-1\right)
}g_{\sigma}\left(  z^{n+1}\right)  ,
\]
and in the limit $n\rightarrow\infty$ or $n=1$, $h_{\sigma}(z)$ returns to the
Bose-Einstein integral $g_{\sigma}(z)$ or the Fermi-Dirac integral $f_{\sigma
}(z)$, respectively \cite{OursAnn}.

We will show that such an ideal gas system may display the BEC type phase
transition, and whether the phase transition occurs or not lies on the value
of $k$, the position of the state with an infinite maximum occupation number
in the spectrum.

\section{The phase transition}

In this section, we consider two cases which can display BEC type phase
transitions: the ideal gases obeying the generalized statistics with
$n_{0}=\infty$, $n_{i}=n$ ($i\neq0$) and with $n_{k}=\infty$ ($k\neq0$),
$n_{i}=n$ ($i\neq k$). An interesting case is $n=1$. In this case, the maximum
occupation number of only one state is the same as that in the Bose case, but
of all other states is the same as that in the Fermi case. We will show that
even such systems in which only one state's maximum occupation number is
infinite can still display BEC type phase transitions.

\subsection{The explicit expression for the fugacity}

For judging whether there is a phase transition or not and determining the
phase transition temperature, we first solve the exact explicit expression for
the fugacity from the equation of state, and, then, analyze the discontinuity
in the derivative of the fugacity.

Based on the homogeneous Riemann-Hilbert problem \cite{Musihailishiweili}, we
can solve the explicit expression for the fugacity from equation
(\ref{Neqofs}) exactly (A brief introduction to the method of the
Riemann-Hilbert problem see Ref. \cite{OurRH}).

For the case of $n_{k}=\infty$ and $n_{i}=n$ ($i\neq k$), introduce a complex
function%
\begin{equation}
\Psi\left(  \zeta\right)  =\frac{N_{\lambda}}{N}h_{\nu/s}\left(  \zeta\right)
+\frac{1}{N}\frac{n+1}{\left(  \zeta^{-1}e^{\beta\varepsilon_{k}}\right)
^{n+1}-1}-1,\label{PSI}%
\end{equation}
where $h_{\sigma}\left(  \zeta\right)  $ is an analytic continuation of
$h_{\sigma}(z)$. From equation (\ref{Neqofs}), we can see that the fugacity is
a zero of $\Psi\left(  \zeta\right)  $ on the real axis. Therefore, the
problem of solving the fugacity $z$ is converted into the problem of seeking
the real zero of the complex function $\Psi\left(  \zeta\right)  $.

We can express $\Psi\left(  \zeta\right)  $ as
\begin{equation}
\Psi\left(  \zeta\right)  =\eta\frac{\left(  \zeta-z\right)
{\displaystyle\prod\limits_{i=1}^{n_{zero}-1}}
\left(  \zeta-\omega_{i}\right)  }{%
{\displaystyle\prod\limits_{j=1}^{n_{\rho}}}
\left(  \zeta-\rho_{j}\right)
{\displaystyle\prod\limits_{m=1}^{n_{b}}}
\left(  \zeta-c_{m}\right)  ^{\kappa_{m}}}\varphi\left(  \zeta\right)
,\label{psiandphi}%
\end{equation}
where $\varphi\left(  \zeta\right)  $ is the fundamental solution of the
homogeneous Riemann-Hilbert problem, which has no zeros and singularities, $z
$ (the fugacity) and $\omega_{i}$ are zeros of $\Psi\left(  \zeta\right)  $,
$n_{zero}$ is the number of the zeros, $\rho_{j}$ is a pole of $\Psi\left(
\zeta\right)  $, $n_{\rho}$ is the number of the poles, $c_{m}$ is an endpoint
that is different from infinity of the boundary of the analytic region of
$\Psi\left(  \zeta\right)  $ (in the present case, the boundary of the
analytic region of $\Psi\left(  \zeta\right)  $ is a set of rays (see figure
1) with the origins $c_{m}$), $n_{b}$ is the number of the endpoints different
from infinity of the boundary (in the present case $n_{b}$ is the number of
the rays), the constant $\kappa_{m}$ is introduced to equal the degrees of
divergence of the two sides of this equation at the $m$-th endpoint $c_{m}$,
and $\eta$ is a constant.

From equation (\ref{psiandphi}), it is easy to see that we can in principle
obtain an explicit expression of the fugacity $z$. For this purpose, we need
to first determine the fundamental solution $\varphi\left(  \zeta\right)  $,
the endpoints $c_{m}$, and the poles $\rho_{j}$, etc.

\textit{The analytic region.} For determining $c_{m}$, the endpoints of the
boundary of the analytic region, we first analyze the analytic region of
$\Psi\left(  \zeta\right)  $.

The boundary of the analytic region of $\Psi\left(  \zeta\right)  $ is
determined by the analytic region of%
\begin{equation}
h_{\sigma}\left(  \zeta\right)  =g_{\sigma}\left(  \zeta\right)  -\frac
{1}{\left(  n+1\right)  ^{\sigma-1}}g_{\sigma}\left(  \zeta^{n+1}\right)  ,
\end{equation}
where $g_{\sigma}\left(  \zeta\right)  $ is the analytically continued
Bose-Einstein integral which is just the Jonqui\'{e}re function \cite{MOS}:%
\[
g_{\sigma}\left(  \zeta\right)  =Li_{\sigma}\left(  \zeta\right)  .
\]
The boundary of the analytic region of $Li_{\sigma}\left(  \zeta\right)  $ is
the positive real axis from $1$ to $\infty$ \cite{MOS}. Consequently, the
boundary of the analytic region of $h_{\sigma}\left(  \zeta\right)  $ and
$\Psi\left(  \zeta\right)  $ consists of $n$ rays with origins on the unit
circle (figure 1), denoted as $L_{m}$, $m=1,2,\cdots,n$, i.e., the $m$-th ray
$L_{m}$ is $\left[  e^{i\frac{2\pi m}{n+1}},\infty e^{i\frac{2\pi m}{n+1}%
}\right)  $. It should be emphasized that $h_{\sigma}\left(  \zeta\right)  $
has no singularity on the positive real axis, or, $h_{\sigma}\left(
\zeta\right)  $ is analytic on the positive real axis. Then, the endpoints
that are different from infinity of the boundary (the origins of the rays) are%
\begin{equation}
c_{m}=e^{i\frac{2\pi m}{n+1}},\text{ \ \ }m=1,2,\cdots,n.
\end{equation}
Therefore we have $n_{b}=$ $n$.

\textit{The fundamental solution of the homogeneous Riemann-Hilbert problem.
}Now, we calculate the fundamental solution of the homogeneous Riemann-Hilbert
problem, $\varphi\left(  \zeta\right)  $.

Using the result of the homogeneous Riemann-Hilbert problem
\cite{Musihailishiweili}, we have%
\begin{equation}
\varphi\left(  \zeta\right)  =e^{\gamma\left(  \zeta\right)  }%
{\displaystyle\prod\limits_{m=1}^{n}}
\left(  \zeta-c_{m}\right)  ^{\lambda_{m}}.\label{phi}%
\end{equation}
Here%
\begin{align}
\gamma\left(  \zeta\right)   &  =\frac{1}{2\pi i}\sum_{m=1}^{n}\int_{L_{m}%
}dx\frac{\ln G\left(  x\right)  }{x-\zeta}\nonumber\\
&  =\frac{1}{2\pi i}\sum_{m=1}^{n}e^{i\frac{2\pi m}{n+1}}\int_{1}^{\infty
}dx\frac{\ln G\left(  xe^{i\frac{2\pi m}{n+1}}\right)  }{xe^{i\frac{2\pi
m}{n+1}}-\zeta},\label{gamma1}%
\end{align}
where%
\begin{equation}
G\left(  \zeta\right)  =\frac{\varphi^{+}\left(  \zeta\right)  }{\varphi
^{-}\left(  \zeta\right)  }%
\end{equation}
is the jump of $\varphi\left(  \zeta\right)  $ on the boundary, and
$\lambda_{m}$ is an integer determined by the condition%
\begin{align}
\mp\operatorname{Re}\frac{\ln G\left(  c_{m}\right)  }{2\pi i}+\lambda_{m}  &
=0,\text{ \ \ if}\mp\operatorname{Re}\frac{\ln G\left(  c_{m}\right)  }{2\pi
i}\text{ is an integer,}\nonumber\\
-1  &  <\mp\operatorname{Re}\frac{\ln G\left(  c_{m}\right)  }{2\pi i}%
+\lambda_{m}<0,\text{ \ otherwise.}\label{lbdk1}%
\end{align}

We first need to analyze the analytic region of the fundamental solution
$\varphi\left(  \zeta\right)  $. From equation (\ref{psiandphi}), we can see
that the boundary of the analytic region of $\varphi\left(  \zeta\right)  $
consists of the non-isolated singularities of $\Psi\left(  \zeta\right)  $,
and the jump of $\varphi\left(  \zeta\right)  $ on the boundary is the same as
that of $\Psi\left(  \zeta\right)  $, i.e.,%
\begin{equation}
G\left(  \zeta\right)  =\frac{\Psi^{+}\left(  \zeta\right)  }{\Psi^{-}\left(
\zeta\right)  }.
\end{equation}

The value of $\Psi\left(  \zeta\right)  $ on the two sides of $L_{m}$ is
determined by the value of the function $h_{\sigma}\left(  \zeta\right)  $ on
the two sides of $L_{m}$:%
\begin{equation}
h_{\sigma}^{\pm}\left(  xe^{i\frac{2\pi m}{n+1}}\right)  =g_{\sigma}\left(
xe^{i\frac{2\pi m}{n+1}}\right)  -\frac{1}{\left(  n+1\right)  ^{\sigma-1}%
}\mathfrak{g}_{\sigma}\left(  x^{n+1}\right)  \mp i\frac{\pi}{\Gamma\left(
\sigma\right)  }\left(  \ln x\right)  ^{\sigma-1},\ \ m=1,2,\cdots,n,
\end{equation}
where $\mathfrak{g}_{\sigma}\left(  \zeta\right)  $ is the Cauchy principal
value of $g_{\sigma}\left(  \zeta\right)  $ on the boundary \cite{CW}. Then£¬
\begin{align}
\Psi^{\pm}\left(  xe^{i\frac{2\pi m}{n+1}}\right)   &  =\frac{N_{\lambda}}%
{N}\left[  g_{\nu/s}\left(  xe^{i\frac{2\pi m}{n+1}}\right)  -\frac{1}{\left(
n+1\right)  ^{\nu/s-1}}\mathfrak{g}_{\nu/s}\left(  x^{n+1}\right)  \right]
\nonumber\\
&  +\frac{1}{N}\frac{n+1}{\left(  x^{-1}e^{\beta\varepsilon_{k}}\right)
^{n+1}-1}-1\mp i\frac{\pi}{\Gamma\left(  \nu/s\right)  }\frac{N_{\lambda}}%
{N}\left(  \ln x\right)  ^{\nu/s-1}.\label{psipn}%
\end{align}
Note that, in the case of $n_{k}=\infty$ and $n_{i}=1$ ($i\neq k$), equation
(\ref{psipn}) reduces to%
\begin{equation}
\Psi^{\pm}\left(  -x\right)  =\frac{N_{\lambda}}{N}\mathfrak{f}_{\nu/s}\left(
-x\right)  +\frac{1}{N}\frac{2}{\left(  x^{-1}e^{\beta\varepsilon_{k}}\right)
^{2}-1}-1\mp i\frac{\pi}{\Gamma\left(  \nu/s\right)  }\frac{N_{\lambda}}%
{N}\left(  \ln x\right)  ^{\nu/s-1},
\end{equation}
where $\mathfrak{f}_{\nu/s}\left(  \xi\right)  $ is the Cauchy principal value
of the analytically continued Fermi-Dirac integral.

Next, we calculate $\lambda_{m}$ from equation (\ref{lbdk1}).

At the endpoints of the boundary (including both the endpoints that are
different from infinity and the infinity), we have%
\begin{equation}
G\left(  e^{i\frac{2\pi m}{n+1}}\right)  =G\left(  \infty\right)  =1.
\end{equation}
Choosing $\ln G\left(  \infty\right)  =0$ gives%
\begin{equation}
\ln G\left(  e^{i\frac{2\pi m}{n+1}}\right)  =i\arg G\left(  e^{i\frac{2\pi
m}{n+1}}\right)  =-i2\pi.
\end{equation}
Then we have
\begin{equation}
\lambda_{m}=-1.
\end{equation}
Therefore, the fundamental solution is%
\begin{equation}
\varphi\left(  \zeta\right)  =e^{\gamma\left(  \zeta\right)  }%
{\displaystyle\prod\limits_{m=1}^{n}}
\frac{1}{\zeta-e^{i\frac{2\pi m}{n+1}}}=e^{\gamma\left(  \zeta\right)  }%
\frac{\zeta-1}{\zeta^{n+1}-1}.
\end{equation}

\textit{The value of }$\kappa_{m}$\textit{.} The parameter $\kappa_{m}$ is
chosen to guarantee the degrees of divergence of the two sides of equation
(\ref{psiandphi}) at the endpoint $c_{m}$ to be the same.

At the origin of $L_{m}$, $c_{m}=e^{i2\pi m/\left(  n+1\right)  }$, when
$\nu/s>1$, $\mathfrak{g}_{\nu/s}\left(  x^{n+1}\right)  $ and $\left(  \ln
x\right)  ^{\nu/s-1}$ are convergent, and when $\nu/s\leq1$, the degrees of
divergence of $\mathfrak{g}_{\nu/s}\left(  x^{n+1}\right)  $ and $\left(  \ln
x\right)  ^{\nu/s-1}$ are less than one. Thus we have%
\begin{equation}
\kappa_{m}=0.
\end{equation}

\textit{The isolated singularity. }$\Psi\left(  \zeta\right)  $ has only one
isolated singularity,%
\begin{equation}
\rho=e^{\beta\varepsilon_{k}}.
\end{equation}

\textit{The number of zeros of }$\Psi\left(  \zeta\right)  $. Substituting the
above result into equation (\ref{psiandphi}) gives%
\begin{equation}
\eta\left(  \zeta-z\right)
{\displaystyle\prod\limits_{i=1}^{n_{zero}-1}}
\left(  \zeta-\omega_{i}\right)  =e^{-\gamma\left(  \zeta\right)  }\frac
{\zeta^{n+1}-1}{\zeta-1}\left(  \zeta-e^{\beta\varepsilon_{k}}\right)
\Psi\left(  \zeta\right)  .\label{eqs}%
\end{equation}
In principle, if $\omega_{i}$ and $\eta$ are known, one can obtain the
explicit expression of $z$ directly. Nevertheless, the difficulty of finding
the zeros $\omega_{i}$ is often the same as the difficulty of finding the zero
$z$. Alternatively, we can construct a set of equations for $z$, $\omega_{i}$,
and $\eta$, and obtain $z$ by solving the equations.

For solving $z$, we need $n_{zero}+1$ equations. Since the number of the
isolated singularities of $\Psi\left(  \zeta\right)  $ is already known, the
number of the zeros, $n_{zero}$, can be determined by the argument principle,
the contour being chosen as in figure 1. The result shows that $\Psi\left(
\zeta\right)  $ has $n+1$ zeros on the $\zeta$-plane, so we need $n+2$
equations for determining $z$.

\begin{figure}[ptb]
\begin{center}
\includegraphics[width=14cm]{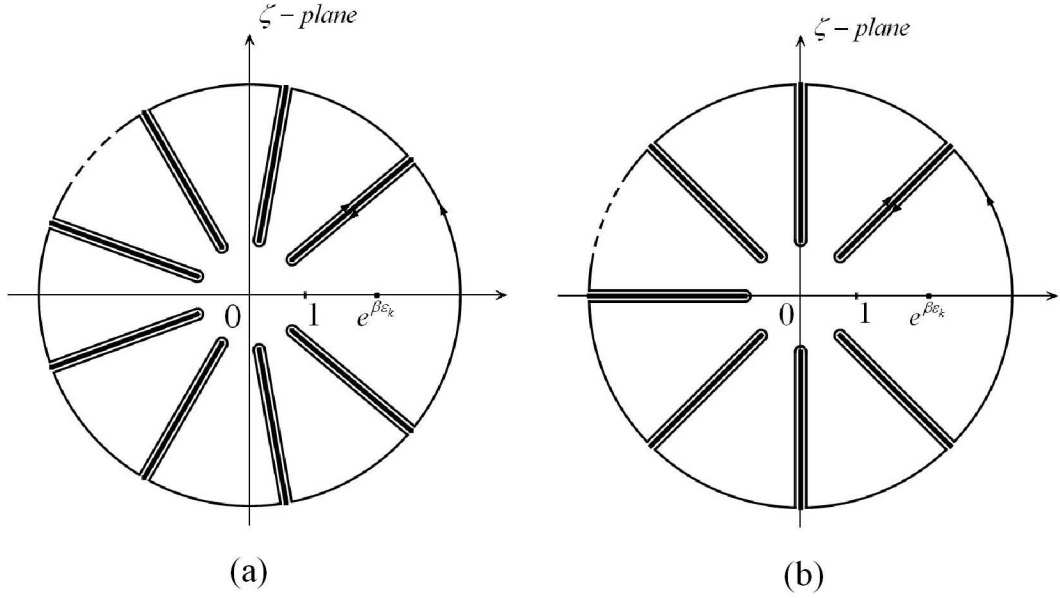}
\end{center}
\caption{The analytic region of $\Psi\left(  \zeta\right)  $. The cases of
$n=even$ and $n=odd$ are illustrated in (a) and (b), respectively.}%
\end{figure}

\textit{The case of }$n_{k}=\infty$\textit{\ and }$n_{i}=1$\textit{\ (}$i\neq
k$\textit{). }For simplicity, we consider the case of $n_{k}=\infty$ and
$n_{i}=1$ ($i\neq k$); the solutions for more general cases can also be
obtained exactly but with more complex forms. When $n=1$, equation (\ref{eqs})
becomes%
\begin{equation}
\eta\left(  \zeta-z\right)  \left(  \zeta-\omega\right)  =e^{-\gamma\left(
\zeta\right)  }\left(  \zeta+1\right)  \left(  \zeta-e^{\beta\varepsilon_{k}%
}\right)  \Psi\left(  \zeta\right)  .\label{eqn1}%
\end{equation}
In this case, for solving $z$, we need three equations.

Substituting $\zeta=0$ into equation (\ref{eqn1}) gives%
\begin{equation}
\eta z\omega=e^{\beta\varepsilon_{k}-\gamma\left(  0\right)  };\label{0order}%
\end{equation}
substituting equation (\ref{0order}) into the derivative of equation
(\ref{eqn1}) and setting $\zeta=0$ give%
\begin{equation}
\gamma^{\prime}\left(  0\right)  -1+e^{-\beta\varepsilon_{k}}-\frac{1}%
{z}-\frac{1}{\omega}=-\frac{N_{\lambda}}{N};\label{1order}%
\end{equation}
substituting $\zeta=e^{\beta\varepsilon_{k}}$ into equation (\ref{eqn1}) gives%
\begin{equation}
\eta\left(  e^{\beta\varepsilon_{k}}-z\right)  \left(  e^{\beta\varepsilon
_{k}}-\omega\right)  =-e^{-\gamma\left(  e^{\beta\varepsilon_{k}}\right)
}\left(  e^{\beta\varepsilon_{k}}+1\right)  \frac{e^{\beta\varepsilon_{k}}}%
{N}.\label{sp}%
\end{equation}
Solving the equations (\ref{0order}), (\ref{1order}) and (\ref{sp}), we have%
\begin{equation}
z=2\left[  \eta_{\lambda}+e^{-\beta\varepsilon_{k}}+\sqrt{\left(
\eta_{\lambda}-e^{-\beta\varepsilon_{k}}\right)  ^{2}+\frac{4e^{-\beta
\varepsilon_{k}}\left(  1+e^{-\beta\varepsilon_{k}}\right)  }{Ne^{\gamma
\left(  e^{\beta\varepsilon_{k}}\right)  -\gamma\left(  0\right)  }}}\right]
^{-1},\label{z}%
\end{equation}
where $\eta_{\lambda}=N_{\lambda}/N+\gamma^{\prime}\left(  0\right)  -1$ and
\begin{equation}
\gamma\left(  \zeta\right)  =\frac{1}{2\pi i}\int_{1}^{\infty}dx\frac{\ln
G\left(  -x\right)  }{x+\zeta}.
\end{equation}

\subsection{The phase transition and the necessary condition for phase
transitions --- the thermodynamic limit}

Equation (\ref{z}) is an exact expression of the fugacity, which
simultaneously describes both gas phase and condensed phase, and, of course,
can describe the transition between these two phases. From equation (\ref{z}),
we can directly see that the thermodynamic limit, $N\rightarrow\infty$, is the
necessary condition for the phase transition.

The fugacity given by equation (\ref{z}) is a smooth function and there is no
singularity. Therefore, there is no phase transition regardless of how low the
temperature is. However, in the thermodynamic limit, i.e., $N\rightarrow
\infty$, equation (\ref{z}) becomes%
\begin{equation}
z=2\frac{1}{\eta_{\lambda}+e^{-\beta\varepsilon_{k}}+\left\vert \eta_{\lambda
}-e^{-\beta\varepsilon_{k}}\right\vert }=\left\{
\begin{array}
[c]{ll}%
e^{\beta\varepsilon_{k}}, & \text{when }\eta_{\lambda}<e^{-\beta
\varepsilon_{k}},\\
\frac{1}{\eta_{\lambda}}, & \text{when }\eta_{\lambda}>e^{-\beta
\varepsilon_{k}}.
\end{array}
\right. \label{zNinf}%
\end{equation}
The discontinuity may appear in the first-order derivative of the fugacity and
the phase transition may occur. The discontinuous point appears at
\begin{equation}
\eta_{\lambda}=e^{-\beta\varepsilon_{k}},\label{Tceq}%
\end{equation}
which is just the phase transition point.

In the phase transition theory, there is a fundamental law: the necessary
condition for a phase transition is that the system must be infinite, i.e.,
the thermodynamic limit. The proof of this statement depends on the assumption
that a finite volume can accommodate at most a finite number of particles. If
the number of particles is finite, the partition function will be an analytic
function and, consequently, there is no singularity in the thermodynamic
function and there is no phase transition \cite{Huang,vanHove,KU}. Such an
assumption is equivalent to requiring that the particle must have a nonzero
volume. Clearly, this assumption does not hold for ideal gases. That is to
say, though this conclusion is valid for all realistic systems (realistic
gases are non-ideal gases), this proof is not valid for the idealized model
--- ideal gases. Our above result shows that for ideal gas systems, the
thermodynamic limit is still a condition for a phase transition.

\subsection{The case of $n_{0}=\infty$ and $n_{i}=1$ ($i\neq0$): the phase
transition temperature}

We first consider the case of $n_{0}=\infty$ and $n_{i}=1$ ($i\neq0$), i.e.,
the state whose maximum occupation number is infinite is the ground state,
$\varepsilon_{k}=\varepsilon_{0}=0$.

In any dimension, there must exist a phase transition. This can be verified
directly by observing the discontinuity in the derivative of the fugacity $z$
from equation (\ref{Tceq}) and the transition temperature is determined by%
\begin{equation}
\eta_{\lambda}=1.\label{Tceq0}%
\end{equation}
Then the phase transition temperature reads%
\begin{equation}
T_{c}=\frac{h^{s}}{2\pi^{s/2}mk_{B}}\left[  \frac{N}{V}\frac{s\Gamma\left(
\nu/2\right)  }{2\Gamma\left(  \nu/s\right)  }\frac{1}{(1-2^{1-\nu/s}%
)\zeta\left(  \nu/s\right)  }\right]  ^{s/\nu}.
\end{equation}

Now let us see what happens when a phase transition occurs. The total number
of the excited particles, from equation (\ref{Neqofs}), is
\begin{equation}
N_{e}=N_{\lambda}f_{\nu/s}\left(  z\right)  .
\end{equation}
Comparing the expressions of $T_{c}$\ and $N_{e}$ gives that when the phase
transition occurs,%
\begin{equation}
N_{e}=N.
\end{equation}
This is just the condition that one determines the phase transition
temperature for a BEC in an ideal Bose gas. In our case, however, this result
comes from a mathematically rigorous calculation rather than being put in by hand.

This result indicates that when the phase transition occurs, the macroscopic
properties of the system will begin to be controlled, to a certain extent, by
a unique quantum state (here the state is the ground state). Such a phase
transition is a sudden change in the particle distribution: in the gas phase,
the macroscopic behavior of the system is a mean contribution of all quantum
states, but in the condensed phase, the quantum state with an infinite maximum
occupation number dominates. This is a BEC type phase transition. More
concretely, when the phase transition begins, the number of excited particles
decreases as the temperature decreases, while the number of particles in the
ground state increases as the temperature decreases:%

\begin{equation}
\frac{N_{e}}{N}=\left(  \frac{T}{T_{c}}\right)  ^{\nu/s},\text{\quad\quad
\quad}\frac{N_{0}}{N}=1-\left(  \frac{T}{T_{c}}\right)  ^{\nu/s}.
\end{equation}

Different from the BEC in an ideal Bose gas, the BEC type phase transition can
occur in any dimension in the ideal gases obeying the statistics in which the
maximum occupation number of the ground state is $\infty$ and of all other
states is finite, since the phase transition can occur for any positive value
of $\nu/s$. In the Bose case, however, the BEC only occurs when $\nu/s>1$,
and, as a result, the BEC only occurs in three-dimensional Bose gases. This is
because the Bose-Einstein integral in the Bose case is replaced by $h_{\nu
/s}\left(  z\right)  $ in the present case, while $h_{\nu/s}\left(  z\right)
$ is always bounded for $0\leq z\leq1$. That is to say, in such an ideal
generalized-statistics gas, the occurrence of the phase transition is easier
than that in a Bose system.

\subsection{The case of $n_{k}=\infty$ and $n_{i}=1$ ($i\neq k$): the phase
transition temperature and the Fermi energy}

Next, we consider the case that the only state with an infinite maximum
occupation number is not the ground state, i.e., $n_{k}=\infty$ and $n_{i}=1$
($k\neq0$ and $i\neq k$). More general cases can be treated by the same procedure.

\textit{The phase transition temperature. }From equation (\ref{zNinf}), we can
see that the discontinuous point of the derivative of the fugacity appears at
$\eta_{\lambda}=e^{-\beta\varepsilon_{k}}$. Equation (\ref{zNinf}) indicates
that the phase transition appears at $z=\omega=e^{\beta\varepsilon_{k}}$.
Substituting $\zeta=e^{\beta\varepsilon_{k}}$ into the derivative of equation
(\ref{eqn1}), when $N\rightarrow\infty$, gives%
\begin{equation}
\Lambda\left(  k_{B}T_{c}\right)  ^{\nu/s}f_{\nu/s}\left(  e^{\frac
{\varepsilon_{k}}{k_{B}T_{c}}}\right)  =1,\label{NeDYN}%
\end{equation}
where $\Lambda=\frac{2\Gamma\left(  \nu/s\right)  }{s\Gamma\left(
\nu/2\right)  }\frac{\left(  2\pi^{s/2}m\right)  ^{\nu/s}}{h^{\nu}}\frac{V}%
{N}$. This result indicates that when the phase transition occurs, the number
of particles in all the states except the infinite-maximum-occupation-number
$k$-th state equals the total number of particles of the system, i.e.,
$N_{n\neq\infty}=N$.

Based on the homogeneous Riemann-Hilbert problem, we can solve the phase
transition temperature from equation (\ref{NeDYN}). For simplicity, we only
give the result for the case of $s=\nu$.

Introduce a complex function%
\begin{equation}
\phi\left(  \tau\right)  =\frac{2}{s\Gamma\left(  s/2\right)  }\frac{V}%
{N}\frac{2\pi^{s/2}m}{h^{s}}k_{B}\tau f_{1}\left(  e^{\frac{\varepsilon_{k}%
}{k_{B}\tau}}\right)  -1.\label{psitau}%
\end{equation}
The phase transition temperature $T_{c}$ is a zero of $\phi\left(
\tau\right)  $ on the real axis.

\textit{The analytic region. }We first analyze the analytic region of
$\phi\left(  \tau\right)  $ on the $\tau$-plane. The analytic region of
$\phi\left(  \tau\right)  $ is determined by the behavior of the analytically
continued Fermi-Dirac integral, $f_{1}\left(  e^{\varepsilon_{k}/\left(
k_{B}\tau\right)  }\right)  $, which is illustrated in figure 2(a). The
boundary of this region is complex. Introducing a transformation%
\begin{equation}
\xi=\frac{1}{k_{B}\tau},
\end{equation}
we have%
\begin{equation}
\psi\left(  \xi\right)  =\frac{2}{s\Gamma\left(  s/2\right)  }\frac{V}{N}%
\frac{2\pi^{s/2}m}{h^{s}}\frac{1}{\xi}f_{1}\left(  e^{\varepsilon_{k}\xi
}\right)  -1.
\end{equation}
\vspace{0.2in}

The boundary of the analytic region of $\psi\left(  \xi\right)  $ on the $\xi
$-plane is%
\begin{align}
\operatorname{Re}\xi &  \geq0,\nonumber\\
\operatorname{Im}\xi &  =\frac{\left(  2q+1\right)  \pi}{\varepsilon_{k}%
},\text{ \ \ \ }q=0,\pm1,\pm2,\cdots,
\end{align}
as illustrated in figure 2(b), which is a set of rays running parallel to the
real axis with origins%
\begin{equation}
c_{q}=\left(  0,\frac{\left(  2q+1\right)  \pi}{\varepsilon_{k}}\right)  .
\end{equation}

\begin{figure}[ptb]
\begin{center}
\includegraphics[width=11cm]{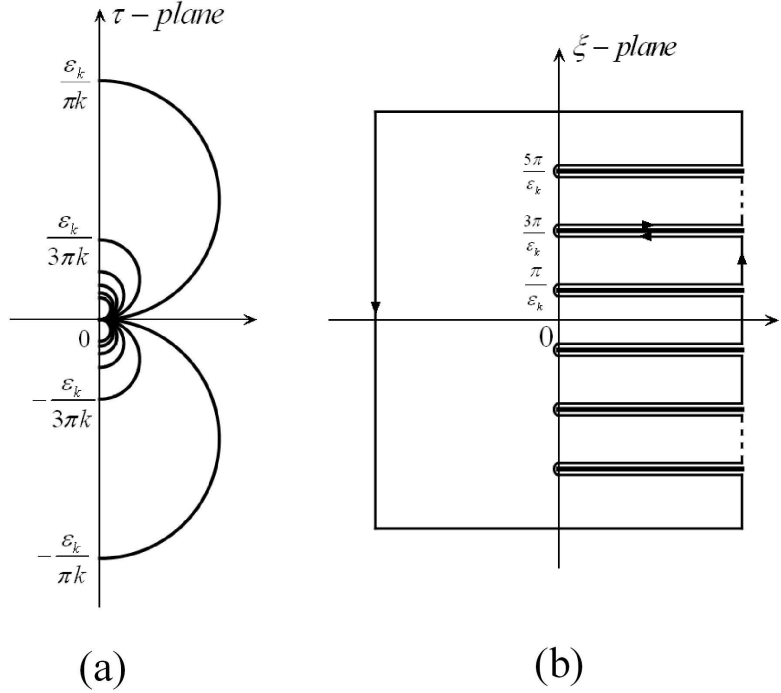}
\end{center}
\caption{(a) The analytic region of $\phi\left(  \tau\right)  $; (b) The
analytic region of $\psi\left(  \xi\right)  $.\ \ \ }%
\end{figure}

\textit{The fundamental solution of the homogeneous Riemann-Hilbert problem.
}We can also express $\psi\left(  \xi\right)  $ in the form of equation
(\ref{psiandphi}), and, then, solve the explicit expression of the phase
transition temperature. First, we seek for the fundamental solution of the
homogeneous Riemann-Hilbert problem. According to equation (\ref{phi}), the
fundamental solution can be written in the following form:%
\begin{equation}
\varphi\left(  \xi\right)  =e^{\gamma\left(  \xi\right)  }%
{\displaystyle\prod\limits_{q=-\infty}^{\infty}}
\left(  \xi-c_{q}\right)  ^{\lambda_{q}},
\end{equation}
where%
\begin{equation}
\gamma\left(  \xi\right)  =\frac{1}{2\pi i}%
{\displaystyle\int\nolimits_{\Sigma_{q}L_{q}}}
d\chi\frac{\ln G\left(  \chi\right)  }{\chi-\xi},
\end{equation}
and the integral is along the boundary of the analytic region,
\begin{equation}
L_{q}:\xi=x+i\frac{\left(  2q+1\right)  \pi}{\varepsilon_{k}},\ x\in\left[
0,\infty\right)  \text{ and }q=0,\pm1,\pm2,\cdots.
\end{equation}
The constant $\lambda_{q}$ is an integer satisfying the condition (\ref{lbdk1}).

The jump on the boundary of the fundamental solution $\varphi\left(
\xi\right)  $ is the same as that of $\psi\left(  \xi\right)  $:%
\begin{equation}
G\left(  \xi\right)  =\frac{\varphi^{+}\left(  \xi\right)  }{\varphi
^{-}\left(  \xi\right)  }=\frac{\psi^{+}\left(  \xi\right)  }{\psi^{-}\left(
\xi\right)  }.\label{Gksi}%
\end{equation}
$\psi^{\pm}\left(  \xi\right)  $, the value of $\psi\left(  \xi\right)  $ on
the two sides of the boundary, is determined by the behavior of the
analytically continued Fermi-Dirac integral,%
\begin{equation}
f_{\sigma}^{\pm}\left(  e^{\left[  x+i\frac{\left(  2q+1\right)  \pi
}{\varepsilon_{k}}\right]  \varepsilon_{k}}\right)  =\mathfrak{f}_{\sigma
}\left(  -e^{x\varepsilon_{k}}\right)  \mp i\frac{\pi}{\Gamma\left(
\sigma\right)  }\left(  x\varepsilon_{k}\right)  ^{\sigma-1}.
\end{equation}
Then,%
\begin{align}
&  \psi^{\pm}\left(  x+i\frac{\left(  2q+1\right)  \pi}{\varepsilon_{k}%
}\right)  =\frac{\Lambda^{\prime}}{x^{2}+\left[  \frac{\left(  2q+1\right)
\pi}{\varepsilon_{k}}\right]  ^{2}}\left\{  x\mathfrak{f}_{1}\left(
-e^{x\varepsilon_{k}}\right)  \mp\pi\frac{\left(  2q+1\right)  \pi
}{\varepsilon_{k}}\right\}  -1\nonumber\\
&  +i\frac{\Lambda^{\prime}}{x^{2}+\left[  \frac{\left(  2q+1\right)  \pi
}{\varepsilon_{k}}\right]  ^{2}}\left\{  \mp\pi x-\mathfrak{f}_{1}\left(
-e^{x\varepsilon_{k}}\right)  \frac{\left(  2q+1\right)  \pi}{\varepsilon_{k}%
}\right\}  ,\label{phitau}%
\end{align}
where $\Lambda^{\prime}=\frac{2}{s\Gamma\left(  s/2\right)  }\frac{2\pi
^{s/2}m}{h^{s}}\frac{V}{N}$. From equations (\ref{Gksi}) and (\ref{phitau}),
we can see that $G\left(  \infty+i\left(  2q+1\right)  \pi/\varepsilon
_{k}\right)  =1$. The constant $\lambda_{q}$ is determined by the condition
(\ref{lbdk1}). Choosing $\ln G\left(  \infty\right)  =0$ gives%
\begin{equation}
\lambda_{q}=0.
\end{equation}
Consequently, the fundamental solution is%
\begin{equation}
\varphi\left(  \xi\right)  =e^{\gamma\left(  \xi\right)  }.
\end{equation}

\textit{The value of }$\kappa_{q}$\textit{.} At the endpoints $c_{q}=\left(
0,\left(  2q+1\right)  \pi/\varepsilon_{k}\right)  $, we have $\psi\left(
i\left(  2q+1\right)  \pi/\varepsilon_{k}\right)  \sim\ln\left[  \xi-i\left(
2q+1\right)  \pi/\varepsilon_{k}\right]  $. Then,%
\begin{equation}
\kappa_{q}=0.
\end{equation}

\textit{The isolated singularity of }$\psi\left(  \xi\right)  $\textit{.
}$\psi\left(  \xi\right)  $ has only one isolated singularity,%
\begin{equation}
\rho=0.
\end{equation}

\textit{The number of the zeros of }$\psi\left(  \xi\right)  $. By the
argument principle, the contour being illustrated in figure 2(b), we can
determine that $\psi\left(  \xi\right)  $ has only one zero, $\xi=\beta
_{c}=1/\left(  k_{B}T_{c}\right)  $, which is on the real axis.

Introducing $\Phi\left(  \xi\right)  =\xi\psi\left(  \xi\right)  $, we have%
\begin{equation}
\Phi\left(  \xi\right)  =\upsilon\left(  \xi-\beta_{c}\right)  e^{\gamma
\left(  \xi\right)  },\label{bigPhi}%
\end{equation}
where $\upsilon$ is a constant. Substituting $\xi=0$ into equation
(\ref{bigPhi}) and its first-order derivative gives two equations. Solving
these equations gives%
\begin{align}
T_{c} &  =\frac{h^{s}}{2\pi^{s/2}mk_{B}}\frac{N}{V}\frac{s}{2}\Gamma\left(
\frac{s}{2}\right)  \frac{1}{\ln2}-\frac{1}{2\ln2}\frac{\varepsilon_{k}}%
{k_{B}}\nonumber\\
&  +\frac{1}{k_{B}}\frac{1}{2\pi i}\sum_{q=-\infty}^{\infty}%
{\displaystyle\int\nolimits_{0}^{\infty}}
dx\frac{\ln G\left(  x+i\left(  2q+1\right)  \pi/\varepsilon_{k}\right)
}{\left(  x+i\left(  2q+1\right)  \pi/\varepsilon_{k}\right)  ^{2}}.\label{Tc}%
\end{align}
The last term of equation (\ref{Tc}) is small when $\varepsilon_{k}$ is small,
and is roughly proportional to $\varepsilon_{k}^{2}$.

The explicit expression of the phase transition temperature shows that $T_{c}
$ depends on the value of $\varepsilon_{k}$. Especially, $T_{c}=0$ appears at%
\begin{equation}
\varepsilon_{k}=\frac{h^{s}}{2\pi^{s/2}m}\left[  \Gamma\left(  \frac{\nu}%
{2}+1\right)  \frac{N}{V}\right]  ^{s/\nu}\equiv\epsilon_{F},
\end{equation}
i.e., when $\varepsilon_{k}\geq\epsilon_{F}$, there will be no phase
transition. Note that this result holds also for the case of $s\neq\nu$. It is
not difficult to recognize the physical meaning of $\epsilon_{F}$: it is just
the Fermi energy of a $\nu$-dimensional ideal Fermi gas with the dispersion
relation $\varepsilon=p^{s}/\left(  2m\right)  $ \cite{OursAnn}. The reason
why there is no phase transition when $\varepsilon_{k}>\epsilon_{F}$ is that
if $\varepsilon_{k}>\epsilon_{F}$, the states below $\epsilon_{F}$ can
accommodate all particles in the system and, then, there are no enough
particles accumulating in the $k$-th state, i.e., the BEC type phase
transition cannot occur.

\section{Discussions and outlook}

In this paper, we construct an exactly solvable phase transition model. We
first consider a generalized statistics in which the maximum occupation
numbers of different quantum states can take on different values. When the
maximum occupation numbers of all the states are the same, e.g., equaling
$\infty$, $1$, or an arbitrary integer, the generalized statistics returns to
Bose-Einstein, Fermi-Dirac, or Gentile statistics
\cite{OursAnn,Gentile,OursPLA}, respectively. The model constructed in this
paper is an ideal gas obeying the generalized statistics in which the maximum
occupation number of only one state is infinite, but of all other states is
finite. The phase transition which occurs in such systems is the BEC type
phase transition. For judging if the phase transition can occur and
determining the phase transition point, we calculate the exact explicit
solution for the fugacity with the help of the mathematical result of the
homogeneous Riemann-Hilbert problem. By observing the discontinuity in the
derivative of the fugacity, we analyze the phase transition rigorously. From
this phase transition model, we can see that the thermodynamic limit is a
necessary condition for a phase transition of an ideal system.

For constructing the solvable phase transition model, we introduce a kind of
intermediate statistics. Various generalized exclusion statistics models play
important roles in many fields
\cite{Gentile,gPauli,Wilczek,Ha,Isakov,Surth,Wadati}, since many physical
systems may behave as neither Bose-Einstein nor Fermi-Dirac system. Though
nature realizes only bosons and fermions, there are many composite-particle
systems, e.g., the Cooper pair in the theory of superconductivity, the Fermi
gas superfluid \cite{FermiSuperfluid}, the exciton \cite{exciton}, the magnon
\cite{OurJSATA}, etc. For example, a boson consists of two fermions obeys
Bose-Einstein statistics. However, when two such bosons come closer together,
the fermions in the composite bosons may "feel" each other, and the statistics
may somewhat deviate from Bose-Einstein statistics. In this case, such a
composite system can be effectively viewed as obeying a kind of intermediate
statistics. It is shown in a resent study \cite{Guan} that the fermion pairs
in the one-dimensional Fermi gases obey generalized exclusion statistics.

The experimental and theoretical research of BEC is a branch of the
statistical physics of a rapidly growing importance \cite{BEC}. The BEC of
ideal Bose gases is a special case of the generalized BEC phase transition. By
studying this exactly solvable model, we can also obtain a deeper insight into
the BEC of ideal Bose gases. We can conclude that the conditions for the BEC
type phase transition as follows:

(1) There must exist a \textit{low enough} quantum state with an infinite
maximum occupation number, where "low enough" means that when the temperature
tends to the absolute zero, there are still a macroscopic number of particles
in this state. In other words, this condition requires that the total capacity
of all states below such a state must be small enough so that this state can
be macroscopically occupied when the temperature is low. As a result, the
energy of this state, denoted as $\varepsilon_{\infty}^{\min}$, must be the
lowest one among the states whose maximum occupation numbers are infinite.
Such a condition is of course satisfied by a Bose system since the maximum
occupation number of the ground state is infinite. However, for the systems
obeying the generalized statistics, as discussed above, in the case of
$n_{0}=\infty$ and $n_{i}=n$ ($i\neq0$), this condition can be always
satisfied, but in more general cases, e.g., the case of $n_{k}=\infty$ and
$n_{i}=n$ ($i\neq k$), this condition can be satisfied only when
$\varepsilon_{k}<\epsilon_{F}$.

(2) The state that will be macroscopically occupied when a BEC type phase
transition occurs must be isolated from other
infinite-maximum-occupation-number states, where "isolated" means that the
state density of this state is a $\delta$-function, i.e., the state density of
the states with infinite maximum occupation numbers (not the state density of
the system) must take the form of $\rho_{\infty}\left(  \varepsilon\right)
=\eta_{\infty}\left(  \varepsilon\right)  +\delta\left(  \varepsilon
-\varepsilon_{\infty}^{\min}\right)  $ and $\eta_{\infty}\left(
\varepsilon_{\infty}^{\min}\right)  =0$, where $\rho_{\infty}\left(
\varepsilon\right)  $ is the density of the infinite-maximum-occupation-number
states. In the examples of the generalized statistics we considered above,
this condition is satisfied naturally, since there is only one
infinite-maximum-occupation-number state, for $n_{0}=\infty$ and $n_{i}=n$
($i\neq0$), $\rho_{\infty}\left(  \varepsilon\right)  =\delta\left(
\varepsilon\right)  $, and for $n_{k}=\infty$ and $n_{i}=n$ ($i\neq k$),
$\rho_{\infty}\left(  \varepsilon\right)  =\delta\left(  \varepsilon
-\varepsilon_{k}\right)  $. However, for the case of ideal Bose gases, this
condition is not always satisfied. In ideal Bose gas systems, the maximum
occupation number of all states is infinite, i.e., the state density of the
system $\rho\left(  \varepsilon\right)  =\rho_{\infty}\left(  \varepsilon
\right)  $, and then the lowest infinite-maximum-occupation-number state is
the ground state, i.e., $\varepsilon_{\infty}^{\min}=0$. In three dimensions,
the state density is $\rho_{\infty}\left(  \varepsilon\right)  =\eta_{\infty
}\left(  \varepsilon\right)  +\delta\left(  \varepsilon\right)  $, where
$\eta_{\infty}\left(  \varepsilon\right)  \propto\sqrt{\varepsilon}$, so
$\eta_{\infty}\left(  0\right)  =0$. The condition is satisfied, and the BEC
phase transition can occur in a three-dimensional ideal Bose gas. In one and
two dimensions, the state densities are $\rho_{\infty}\left(  \varepsilon
\right)  =\eta_{\infty}\left(  \varepsilon\right)  \propto1/\sqrt{\varepsilon
}$ and $\rho_{\infty}\left(  \varepsilon\right)  =\eta_{\infty}\left(
\varepsilon\right)  =const$, respectively; the above condition is not
satisfied, and there are no BEC phase transitions in one- and two-dimensional
Bose gases.

Furthermore, many physical systems possess other kinds of statistics beyond
Bose-Einstein and Fermi-Dirac statistics. For example, the Calogero-Sutherland
model is shown to possess fractional statistics \cite{Ha}, a spinless fermions
system in two dimensions may obey exclusion statistics \cite{Surth}, and bound
pairs of fermions form hard-core bosons obeying generalized exclusion
statistics \cite{Guan}. Moreover, in the model constructed in the present
paper, there are both bosonic and fermionic states in a system. In a Bose
system, if each boson consists of two fermions, then in the system there must
simultaneously exist both bosons and fermions due to the fact that there
exists an "ionization" energy. As long as the temperature of the system is not
the absolute zero, there are always a certain proportion of particles having
energies larger than\ the "ionization" energy and behaving as fermions. In
such a case, the particle at the low-lying state behaves as a boson and the
particle at the high-lying state behaves as a fermion. That is to say, a
composite system will not accurately possess Bose-Einstein or Fermi-Dirac
statistics. In such a composite system, our model may work. We will address
this issue in future work. Moreover, a system consisted of both bosons and
fermions also has been studied in literature \cite{Kitaev}.

\vskip 0.5cm \noindent\textbf{Acknowledgements }

We are very indebted to Dr. G. Zeitrauman for his encouragement. This work is
supported in part by NSF of China under Grant No. 10605013 and the Hi-Tech
Research and Development Programme of China under Grant No. 2006AA03Z407.

\end{document}